\documentclass[10pt]{amsart}

\usepackage[small]{caption}
\usepackage{amsmath,amsfonts, amssymb,epsf, epsfig}\usepackage{graphicx}

\newcommand{\pa}{\partial}

\begin{document}

\title{Repulsive Casimir force   at zero and finite temperature}
\author{S.C. Lim}\address{ Faculty of Engineering,
Multimedia University, Jalan Multimedia, Cyberjaya, 63100, Selangor
Darul Ehsan, Malaysia.}\email{sclim@mmu.edu.my}
\author{L.P. Teo}\address{Faculty of Information
Technology, Multimedia University, Jalan Multimedia, Cyberjaya,
63100, Selangor Darul Ehsan, Malaysia.}\email{lpteo@mmu.edu.my}


\begin{abstract}
 We study the zero and finite temperature Casimir force acting on a perfectly conducting piston with arbitrary cross section moving inside a closed cylinder with infinitely permeable walls. We show that at any temperature, the Casimir force always tends to move the piston away from the walls and towards its equilibrium position. In the case of rectangular piston, exact expressions for the Casimir force are derived. In the high temperature regime, we show that the leading term of the Casimir force is linear in temperature and therefore the Casimir force has a classical limit. Due to duality, all these result also hold for an infinitely permeable piston moving inside a closed cylinder with perfectly conducting walls.

\vspace{0.2cm}\noindent
\end{abstract}
\maketitle
\section{Introduction}\label{sec1}
\begin{figure}\centering \epsfxsize=.6\linewidth
\epsffile{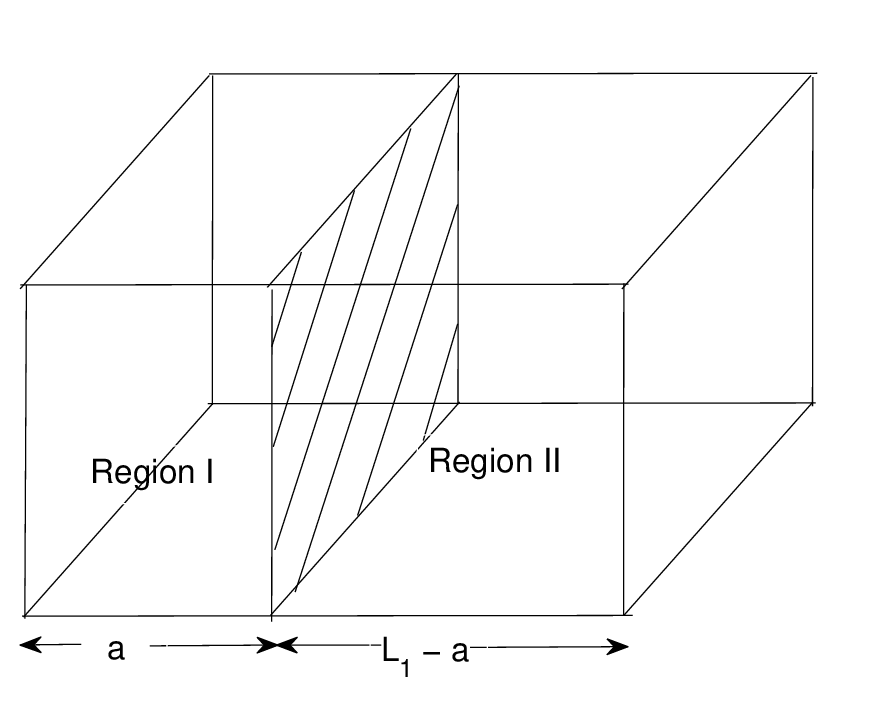} \caption{\label{f1}A rectangular piston moving freely inside a rectangular cylinder dividing the cylinder into Regions I and II.}\end{figure}

It is well known that the vacuum fluctuations of electromagnetic fields in the presence of boundaries give rise to  Casimir force, which was shown to be attractive when the boundary consists of a pair of perfectly conducting parallel plates \cite{1}. Since the seminal work of Casimir, many studies have been done on Casimir effect where different quantum fields and geometric configurations have been considered.  Nevertheless, after 60 years of its discovery, Casimir effect is still under active research. By definition, the calculation of Casimir energy involves an infinite sum that needs to be regularized. However, there is still no consensus on the regularization procedures, which often lead to inconsistent results. Even for   geometric configuration as simple as a rectangular cavity, despite zeta regularization technique or dimensional regularization method can give  finite   results for the Casimir force acting on a wall \cite{2,3}, some authors considered these regularization methods which renormalize all surface divergence terms to zero as being not physical \cite{4}. Nonetheless, Geyer,  Klimchitskaya and   Mostepanenko \cite{nn1} have recently developed a subtraction scheme to obtain a physically consistent Casimir force acting on a wall of a perfectly conducting rectangular cavity from the point of view of thermodynamics.
Another approach to this problem was considered by Fulling et al \cite{nn2}.
In 2004, Cavalcanti \cite{5} proposed an alternative to this problem by adding a piston in the rectangular cavity. He showed that for a 2-dimensional rectangular piston, the Casimir force acting on the piston due to fluctuations of a scalar field with Dirichlet boundary conditions is finite without renormalization and can be computed exactly. Since then, Casimir piston has attracted considerable interest \cite{6,7,8,9,10,11,12,13,14,15,16,17,18,19}. It has been shown that at any dimension and temperature, for either scalar field or electromagnetic field, if the boundary conditions assumed on all the walls are the same and is an ideal condition, then the Casimir force is always an attractive force which tends to pull the piston to the nearer wall. As the length scale shrinks to the nano range, this attractive Casimir force would create undesirable effects on microelectromechanical and nanoelectromechanical devices known as stiction \cite{20, 21}, which would limit the functionality of the devices. As a result, there is an impelling need to search for scenarios that would lead to repulsive Casimir force. In the case of infinite parallel plates,    Boyer \cite{22} has shown that the Casimir force acting on the plates is repulsive if one of the plates is perfectly conducting and the other is infinitely permeable.  The thermal correction of the Casimir force for this   configuration was considered in \cite{23, 24} and it was proved that the Casimir force remains repulsive at any finite temperature. Another generalization of Boyer's work was considered in \cite{25}, where it was proved that if the plates are made of dielectric materials with nontrivial magnetic permeability, then for a large range of parameters, the Casimir force is repulsive. Similar results were obtained in \cite{n2}.   For the   piston geometry,  Barton \cite{9}  showed that the Casimir force acting on a piston made of weakly dielectric materials can become repulsive when the plate separation is sufficiently large. In this paper, we generalize the original setup of Boyer \cite{22} to piston moving freely inside a closed cylinder, where the piston is perfectly conducting and the walls of the cylinder are infinitely permeable (see FIG. \ref{f1}). This setup has been suggested by  Fulling et al \cite{n1} but they only considered the zero temperature Casimir force for rectangular piston and exact explicit formulas of the Casimir force were not given. In the present article, we allow the piston   to have arbitrary cross section and it is  shown that  the Casimir force is free of surface divergence even without renormalization. We prove rigourously that at any temperature, the Casimir force is always repulsive tending to restore the piston to its equilibrium position, and the magnitude of the Casimir force decreases as the piston moves towards the equilibrium position. In the high temperature regime, we show that the Casimir forces due to the blackbody radiation from the two regions separated by the piston cancel each other, and the leading order of the Casimir force is linear in temperature. This implies the existence of classical limit for the Casimir force. Due to the duality between  electric field and magnetic field in (3+1) dimension, all the results in this paper also hold for an infinitely permeable piston moving freely in a closed perfectly conducting cylinder.

In this paper,  we use the units where $\hbar=c=k_B=1$ in sections \ref{sec2} and \ref{sec3}. In sections \ref{sec4} and \ref{sec5}, we use SI units.
\section{Cut-off dependent Casimir energy}\label{sec2}

In this section, we first discuss the eigenfrequencies of the electromagnetic field for  each region divided by the piston. We then give the expressions for the Casimir energy which was computed by exponential cut-off method in Appendix \ref{app}.

\subsection{Modes of the electromagnetic field} For a cylinder $[0, L]\times \Omega$, where the boundary surfaces $\{0\}\times \Omega$ and $\pa\Omega$ are infinitely permeable but the boundary surface $\{L\}\times \Omega$ is perfectly conducting, the eigenfrequencies for the electromagnetic field can be written as follows. The boundary conditions are equivalent to the conditions $\mathbf{n}.\mathbf{E}=0$ and $\mathbf{n}\times \mathbf{B}=\mathbf{0}$ on the surfaces $\{0\}\times \Omega$ and $[0, L]\times \pa\Omega$, and $\mathbf{n}\times\mathbf{E}=\mathbf{0}$ and $\mathbf{n}. \mathbf{B}=0$ on the surface $\{L\}\times \Omega$. Here $\mathbf{n}$ is the unit normal vector to the surface, $\mathbf{E}$ and $\mathbf{B}$ are the electric field   and magnetic field respectively. The fields can be expressed in terms of the vector potential $\mathbf{A}$ by $\mathbf{E}= -\frac{\pa \mathbf{A}}{\pa t}$ and $\mathbf{B}=\nabla\times\mathbf{A}$. Imposing the transversality condition $\text{div} \;\mathbf{A}=0$, Maxwell's equations are equivalent to $\left(\frac{\pa^2}{\pa t^2}-\Delta\right)\mathbf{A}=\mathbf{0}$. One can then check that for the TE modes (i.e. modes with $E_1=0$), a set of independent solutions for $\mathbf{A}$ is given by $A_1=0$,
\begin{equation*}
\begin{split}
A_2= - \cos\frac{\pi\left(k+\frac{1}{2}\right)x_1}{L}\frac{1}{\omega_{D,j}^2}\frac{\pa \phi_{j}(x_2, x_3)}{\pa x_3}e^{i\omega_{\text{TE},j,k} t},\\
A_3=  \cos\frac{\pi\left(k+\frac{1}{2}\right)x_1}{L}\frac{1}{\omega_{D,j}^2}\frac{\pa \phi_{j}(x_2, x_3)}{\pa x_2}e^{i\omega_{\text{TE},j,k} t}.
\end{split}
\end{equation*}Here $k\in \hat{\mathbb{N}}=\mathbb{N}\cup\{0\}$. For $j=1,2,3,\ldots$, $\phi_j(x_2, x_3)$ is a nonzero eigenfunction of the Laplace operator $-\frac{\pa^2}{\pa x_2}-\frac{\pa^2}{\pa x_3^2}$ on $\Omega$ with eigenvalue $\omega_{D,j}^2>0$ and with Dirichlet boundary conditions, i.e. $\left.\phi_j\right|_{\pa\Omega}=0$. The eigenfrequency $\omega_{\text{TE},j,k}$ is given by \begin{equation}\label{eq1}\omega_{\text{TE},j,k}^2=\left(\frac{\pi\left(k+\frac{1}{2}\right)}{L}\right)^2+ \omega_{D,j}^2.\end{equation}
For TM modes (i.e. modes with $B_1=0$), a set of independent solutions for  $\mathbf{A}$ is given by
\begin{equation*}
\begin{split}A_1=&\sin\frac{\pi\left(k+\frac{1}{2}\right)x_1}{L} \psi(x_2, x_3)e^{i\omega_{\text{TM},j,k} t},\\
A_2= & \frac{\pi\left(k+\frac{1}{2}\right)}{L}\cos\frac{\pi\left(k+\frac{1}{2}\right)x_1}{L}\frac{1}{\omega_{N,j}^2}\frac{\pa \psi_j(x_2, x_3)}{\pa x_2}e^{i\omega_{\text{TM},j,k} t},\\
A_3= &\frac{\pi\left(k+\frac{1}{2}\right)}{L}\cos\frac{\pi\left(k+\frac{1}{2}\right)x_1}{L}\frac{1}{\omega_{N,j}^2}\frac{\pa \psi_j(x_2, x_3)}{\pa x_3}e^{i\omega_{\text{TM},j,k} t}.
\end{split}
\end{equation*}Here $k\in \hat{\mathbb{N}}$. For $j=1,2,3,\ldots$, $\psi_j(x_2, x_3)$ is a nonconstant eigenfunction of the Laplace operator  on $\Omega$ with eigenvalue $\omega_{N,j}^2>0$ and with Neumann boundary conditions, i.e. $\left.\frac{\pa\psi_j}{\pa\mathbf{n}}\right|_{\pa\Omega}=0$. $\omega_{\text{TM}, j,k}$ satisfies an equation analogous to \eqref{eq1}.

In the case the piston has rectangular cross section, i.e., $\Omega=[0, L_2]\times [0, L_3]$, we have explicitly
\begin{equation*}\begin{split}
\omega_{D, \boldsymbol{j}}^2=& \left(\frac{\pi j_2}{L_2}\right)^2+\left( \frac{\pi j_3}{L_3}\right)^2, \hspace{0.5cm} \boldsymbol{j}=(j_2, j_3) \in \mathbb{N}^2,\\
\omega_{N, \boldsymbol{j}}^2=& \left(\frac{\pi j_2}{L_2}\right)^2+\left( \frac{\pi j_3}{L_3}\right)^2, \hspace{0.5cm} \boldsymbol{j}=(j_2, j_3) \in \hat{\mathbb{N}}^2\setminus\{\mathbf{0}\}.\end{split}
\end{equation*}
\subsection{Cut-off dependent Casimir energy}
The Casimir energy of the piston system, denoted by $E_{\text{Cas}}^{\text{piston}}(a; L_1)$ is a sum of the Casimir energies of Regions I and II, and the Casimir energy of the   region outside the cylinder, i.e. \begin{equation}\label{eq10_6_1}E_{\text{Cas}}^{\text{piston}}(a; L_1)=E_{\text{Cas}}^{\text{cylinder}}(a)+E_{\text{Cas}}^{\text{cylinder}}(L_1-a)+E_{\text{Cas}}^{\text{out}}.\end{equation} Since the Casimir energy of the exterior region $E_{\text{Cas}}^{\text{out}}$ does not contribute to the Casimir force acting on the piston, we need not compute it here. By definition, the finite temperature Casimir energy of the cylinder $E_{\text{Cas}}^{\text{cylinder}}(L)$ can be written in terms of the partition function $\mathcal{Z}$ associated with a canonical ensemble in the following way:
\begin{equation}\label{eq11_11_1}
E_{\text{Cas}}^{\text{cylinder}}(L)=-T\log \mathcal{Z}= -T\log \prod \frac{e^{-\frac{\omega}{2T}}}{1-e^{-\frac{\omega}{T}}}=\frac{1}{2}\sum \omega   + T\sum \log\left(1-e^{-\omega/T}\right).
\end{equation}Here $\omega$ runs through the set of eigenfrequencies. The first summation on the right hand side of \eqref{eq11_11_1} is divergent. Therefore we
 define a  cut-off dependent Casimir energy $E_{\text{Cas}}^{\text{cylinder}}(L)$ by
\begin{equation}\label{eq10_3_1}
E_{\text{Cas}}^{\text{cylinder}}(L)=\frac{1}{2}\sum \omega e^{-\lambda \omega} + T\sum \log\left(1-e^{-\omega/T}\right).
\end{equation}According to the previous subsection, we can decompose this into the sum over TE modes $E_{\text{Cas, TE}}^{\text{cylinder}}(L)$ and the sum over TM modes $E_{\text{Cas, TM}}^{\text{cylinder}}(L)$. In the appendix, we use zeta function and heat kernel techniques to compute the Casimir energies $E_{\text{Cas, TE}}^{\text{cylinder}}(L)$ and $E_{\text{Cas, TM}}^{\text{cylinder}}(L)$ due to TE modes and TM modes respectively. We find that at zero temperature $T=0$, the  contribution to the cut-off dependent   Casimir energy  of the cylinder $[0, L]\times\Omega$ from the TE modes is given by
\begin{equation}\label{eq9_26_1_1}
\begin{split}
E_{\text{Cas, TE}}^{\text{cylinder}, T=0}(L)=&\frac{6L}{\pi}c_{D,0}\lambda^{-4}+\frac{L}{\sqrt{\pi}}c_{D,1}\lambda^{-3}+\frac{L}{2\pi}c_{D,2}\lambda^{-2}+\frac{L}{8\pi}\left(2\log\lambda +\gamma+2-2\log 2\right)c_{D,4}\\&-\frac{L}{8\pi}\text{FP}_{s=-1}\left(\Gamma(s)\zeta_{\Omega,D}(s)\right)-\frac{1}{2\pi}\sum_{k=1}^{\infty}(-1)^{k}\sum_{j=1}^{\infty}\frac{\omega_{D,j}}{k}K_1\left(2kL\omega_{D,j}\right).
\end{split}
\end{equation}Here $c_{D, j}$, $j=0, 1, 2,\ldots$ are the heat kernel coefficients of the Laplace operator with Dirichlet boundary coefficients on $\Omega$. $\gamma$ is the Euler constant, $K_{\nu}(z)$ is the modified Bessel function and $\text{FP}_{s=s_0}\left(h(s)\right)$ denotes the finite part of the function $h(s)$ at $s=s_0$. Notice that the first five terms of \eqref{eq9_26_1_1} which contain the $\lambda\rightarrow 0^+$ divergent parts of the Casimir energy depend on $L$ linearly. The last term tends to $0$ as $L\rightarrow \infty$.

For the finite temperature Casimir energy, we find that the cut-off dependent Casimir energy \eqref{eq10_3_1} of the cylinder $[0, L]\times\Omega$ due to TE modes is given by
\begin{equation}\label{eq9_26_2}
\begin{split}
E_{\text{Cas, TE}}^{\text{cylinder}}(L)=&\frac{6L}{\pi}c_{D,0}\lambda^{-4}+\frac{L}{\sqrt{\pi}}c_{D,1}\lambda^{-3}+\frac{L}{2\pi}c_{D,2}\lambda^{-2}
+\frac{L}{8\pi}\left(2\log\lambda  +2+\gamma-2\log 2\right)c_{D,4} \\&-\frac{L}{8\pi} \text{FP}_{s=-1}\left(\Gamma(s)\zeta_{\Omega,D}(s)\right)-\frac{LT}{\pi}\sum_{l=1}^{\infty}\sum_{j=1}^{\infty} \frac{\omega_{D,j}}{l}K_1\left(\frac{l\omega_{D,j}}{T}\right)\\&+\frac{T}{2}\sum_{l=-\infty}^{\infty}\sum_{j=1}^{\infty} \log\left(1+e^{-2L\sqrt{(2\pi lT)^2+\omega_{D,j}^2}}\right).
\end{split}
\end{equation}The expressions for the TM modes contributions are similar with TE replaced by TM and $D$ replaced by $N$, where now $c_{N, j}$, $j=1, 2, 3, \ldots$ are the heat kernel coefficients of the Laplace operator   on $\Omega$ with Neumann boundary coefficients, deleting the zero eigenvalue corresponding to constant functions. Analogous to \eqref{eq9_26_1_1}, the first six terms of \eqref{eq9_26_2} depend on $L$ linearly and the last term tends to $0$ as $L\rightarrow \infty$.

In general, the heat kernel coefficients $c_{D/N,0}$, $c_{D/N, 1}$ and $c_{D/N, 2}$ are known to be given by \cite{26}:
\begin{equation*}
c_{D/N, 0}=\frac{A(\Omega)}{4\pi}, \;\;c_{D/N, 1}=\mp\frac{s(\pa\Omega)}{8\sqrt{\pi}}, \;\;c_{D/N, 2}=\chi(\pa\Omega)-\delta_{D/N},
\end{equation*}where $A(\Omega)$ is the area of $\Omega$, $s(\pa\Omega)$ is the arc length of the boundary of $\Omega$ and \begin{equation*}
\chi(\pa\Omega)=\sum_{i}\frac{1}{24}\left(\frac{\pi}{\alpha_i}-\frac{\alpha_i}{\pi}\right)+\sum_{j}\frac{1}{12\pi}\int_{\gamma_j}\kappa(\gamma_j)d\gamma_j,
\end{equation*}with $\alpha_i$ the interior angle of each sharp corner of $\pa\Omega$ and $\kappa(\gamma_j)$  the curvature of each smooth section of $\pa\Omega$. For $l\geq 3$, $c_{D/N, l}$ can be expressed as integrals of functions that depend on the extrinsic and intrinsic curvatures of the boundary $\pa\Omega$ and the boundary conditions imposed. In the special case where $\Omega$ is a rectangle $ [0, L_2]\times[0, L_3]$, we have explicitly
\begin{equation*}
c_{D/N, 0}=\frac{L_2L_3}{4\pi}, \;c_{D/N, 1}=\mp\frac{L_2+L_3}{4\sqrt{\pi}}, \; c_{D/N,2}=\frac{1}{4}-\delta_{D/N},
\end{equation*}and $c_{D/N, l}=0$ for all $l\geq 3$. Here $\delta_D=0$ and $\delta_N=1$ since we exclude the zero mode from the Neumann spectrum.

\section{The Casimir force acting on the piston}\label{sec3}

As shown by \eqref{eq9_26_1_1} or \eqref{eq9_26_2}, the $\lambda\rightarrow 0^+$ divergent terms of the Casimir energy depend linearly on $L$. This implies that $\lambda\rightarrow 0^+$ divergent term of the Casimir energy of the piston system \eqref{eq10_6_1} does not depend on the piston position $a$. Therefore these divergences do not contribute to the Casimir force acting on the piston, which implies that the Casimir force acting on the piston is finite even without renormalization.

\subsection{Finite  temperature   Casimir force and its classical limit}  From \eqref{eq9_26_2}, we find that the Casimir force is given by
\begin{equation}\label{eq10_3_3}\begin{split}
F_{\text{Cas}}(a; L_1) =& -\frac{\pa}{\pa a}E_{\text{Cas}}^{\text{piston}}(a; L_1)
=F_{\text{Cas}}^{ \infty}(a) -F_{\text{Cas}}^{  \infty}(L_1-a),
\end{split}\end{equation} where $F_{\text{Cas}}^{ \infty}(a)$ is the limit of the Casimir force when $L_1\rightarrow \infty$ given by
\begin{equation}\label{eq9_26_3}\begin{split}F_{\text{Cas}}^{ \infty}(a) =T\sum_{l=-\infty}^{\infty}\sum_{ \omega_{D,j}, \omega_{N,j}} \frac{\sqrt{(2\pi lT)^2+\omega^2}}{e^{2a\sqrt{(2\pi lT)^2+\omega^2}}+1}.
\end{split}\end{equation}Since this is obviously a positive decreasing function of $a$, \eqref{eq10_3_3} shows that the Casimir force acting on the piston has positive sign when $a<L_1/2$ and has negative sign when $a>L_1/2$. In other words, the Casimir force always tends to restore the piston to the equilibrium position $a=L_1/2$. Moreover, the magnitude of the Casimir force decreases as the piston moves towards its equilibrium position.

Note that \eqref{eq9_26_3} is an exact expression for the Casimir force at any finite temperature. With the knowledge of the eigenvalues of Laplace operator with Dirichlet and Neumann boundary conditions on the surface $\Omega$, one can compute the Casimir force by this formula to any degree of accuracy. In particular, for a rectangular piston with cross section $[0, L_2]\times[0, L_3]$, we have
\begin{equation*}
\begin{split}F_{\text{Cas}}^{ \infty}(a; L_2, L_3)=&2\pi T\sum_{l=-\infty}^{\infty}\sum_{k_2=1}^{\infty}\sum_{k_3=1}^{\infty}  \frac{\sqrt{(2lT)^2+\left(\frac{k_2}{L_2}\right)^2+\left(\frac{k_3}{L_3}\right)^2}}{\exp\left(2\pi a\sqrt{(2lT)^2+\left(\frac{k_2}{L_2}\right)^2+\left(\frac{k_3}{L_3}\right)^2}\right)+1}\\&+\pi T\sum_{l=-\infty}^{\infty}\sum_{j=2,3}\sum_{k_j=1}^{\infty} \frac{\sqrt{(2lT)^2+\left(\frac{k_j}{L_j}\right)^2}}{\exp\left(2\pi a\sqrt{(2lT)^2+\left(\frac{k_j}{L_j}\right)^2}\right)+1}.
\end{split}
\end{equation*}

Eq. \eqref{eq9_26_3} shows that when $T\rightarrow \infty$, the    Casimir force is dominated by a term linear in $T$ corresponding to those terms with $l=0$. The remaining terms decay exponentially as $T\rightarrow \infty$. Notice that to restore the constants $\hbar, c$ and $k_B$ into the expression for the Casimir force, we need to replace $T$ by $k_BT/(\hbar c)$ everywhere and multiply the overall expression by $\hbar c$. Therefore the high temperature expansion of the Casimir force is the same as the small-$\hbar$ expansion of the Casimir force. A term with order $T^j$ will  be accompanied with the term $\hbar^{1-j}$.  The leading term of the Casimir force \eqref{eq9_26_3} is linear in $T$ implies that this leading term is independent of $\hbar$, and the remaining terms go to zero if we formally let $\hbar$ goes to $0$. As a result, we find that the Casimir force has a classical   limit (or high temperature limit) given by
\begin{equation}\label{eq10_8_6}
F_{\text{Cas}}^{\text{classical}}(a; L_1)=T\sum_{ \omega_{D,j}, \omega_{N,j}}\left\{ \frac{\omega}{e^{2a\omega}+1}-\frac{\omega}{e^{2(L_1-a)\omega}+1}\right\}.
\end{equation}
For its small $a$-behavior, we derive in   Appendix \ref{app} that
\begin{equation*}\begin{split}
F_{\text{Cas}}^{\text{classical}}(a; L_1)
=&\frac{3\zeta_R(3)}{16 \pi a^3}A(\Omega)T+O(a^0).\end{split}
\end{equation*}This shows that at high temperature, when the plate separation $a$ is small, the Casimir force is dominated by the term
\begin{equation*}
F_{\text{Cas}}^{\text{classical}}(a; L_1)\simeq \frac{3\zeta_R(3)}{16 \pi a^3}A(\Omega)T=\frac{3}{4} \frac{\zeta_R(3)}{4 \pi a^3}A(\Omega)T.
\end{equation*}Interestingly, this term does not depend on the geometry of the cross section, but    depends only on the area of the cross section. Moreover, it is equal to $-3/4$ times the classical term of the Casimir force acting on a pair of perfectly conducting parallel plates.
In fact, one can verify that if $\tilde{F}_{\text{Cas}}(a; L_1)$ is the   Casimir force when both the piston and the surrounding   walls are perfectly conducting or infinitely permeable, then the Casimir force $F_{\text{Cas}}(a; L_1)$ when the piston is perfectly conducting but the surrounding walls are infinitely permeable is related to $\tilde{F}_{\text{Cas}}(a; L_1)$ by
\begin{equation*}\begin{split}
&F_{\text{Cas}}(a; L_1)=2\tilde{F}_{\text{Cas}}(2a; 2L_1)-\tilde{F}_{\text{Cas}}(a; L_1).\end{split}
\end{equation*}In particular, when the piston has rectangular cross section, we derive from the results in \cite{18} the following  alternative explicit formula for the classical term of the Casimir force:
\begin{equation*}
\begin{split}&F_{\text{Cas}}^{\text{classical}}(a; L_1, L_2, L_3) = T\Biggl\{\frac{3\zeta_R(3)}{16\pi}\frac{L_2L_3}{a^3}-\frac{\zeta_R(3)}{8\pi}\frac{L_3}{L_2}-\frac{\pi}{24L_3}-\frac{1}{L_2}\sum_{k_2=1}^{\infty}
\sum_{k_3=1}^{\infty}
\frac{k_2}{k_3} \\& \hspace{3cm}\times K_1\left(\frac{2\pi k_2k_3L_3}{L_2}\right) +\frac{\pi L_2L_3}{a^3} \sum_{k_1=1}^{\infty}\sum_{(k_2, k_3)\in \mathbb{Z}^2\setminus\{\textbf{0}\}}\left(k_1+\frac{1}{2}\right)^2 \\& \hspace{3cm} \times K_0\left(\frac{\pi (2 k_1+1)}{a}\sqrt{(k_2L_2)^2+(k_3L_3)^2}\right)\Biggr\} -\left(a\leftrightarrow L_1-a\right).
\end{split}
\end{equation*}

We would like to remark that comparing \eqref{eq9_26_4_a} and \eqref{eq9_26_5_a} in Appendix \ref{app}, we can deduce that in the high temperature regime, the leading behavior of the Casimir force contribution from Region I alone is given by (see Appendix \ref{app}):
\begin{equation}\label{eq10_3_2}\begin{split}
&\frac{T}{\pi}\sum_{l=1}^{\infty}\sum_{j=1}^{\infty} \frac{\omega_{D,j}}{l}K_1\left(\frac{l\omega_{D,j}}{T}\right)+\left(D\leftrightarrow N\right)\\=&\frac{2\pi^3}{45}c_{0,D}T^4+\frac{\zeta_R(3)}{\sqrt{\pi}}c_{1,D}T^3+\frac{\pi}{6}c_{2,D}T^2+O(T)+\left(D\leftrightarrow N\right).
\end{split}\end{equation}This implies that as expected, the leading term of the Casimir force from Region I alone is the blackbody radiation term $$\frac{\pi^2 }{45}A(\Omega)T^4.$$However, since the blackbody radiation from Region II would contribute a force of the same magnitude but opposite direction, the effect of the blackbody radiation on the piston cannot be observed. In fact, there is also a term proportional to $T^2$ of magnitude $$\frac{\pi}{6}(2\chi-1)T^2$$ which has been canceled out in the final Casimir force acting on the piston.

\subsection{Zero temperature Casimir force}In the zero temperature limit, the Casimir force can be obtained  from \eqref{eq9_26_1_1}. When $L_1\rightarrow \infty$, it is given by
\begin{equation*}\begin{split}F_{\text{Cas}}^{ \infty, T=0}(a)=&-\frac{1}{2\pi a}\sum_{k=1}^{\infty}(-1)^{k}\sum_{ \omega_{D,j}, \omega_{N,j}} \frac{\omega }{k}K_1\left(2ka\omega \right)-\frac{1}{\pi }\sum_{k=1}^{\infty}(-1)^{k}\sum_{ \omega_{D,j}, \omega_{N,j}}  \omega^2 K_0\left(2ka\omega \right).
\end{split}\end{equation*}Using the same method as we derive \eqref{eq10_3_2}, one  finds that as $a\rightarrow 0^+$, the leading behavior of the zero temperature Casimir force acting on the piston is given by
\begin{equation}\label{eq10_6_2}\begin{split} F_{\text{Cas}}^{ \infty, T=0}(a)=& \frac{7\pi^3 }{960}\frac{c_{0,D}+c_{0,N}}{a^4}+\frac{3\zeta_R(3)}{16\sqrt{\pi}}\frac{c_{1,D}+c_{1,N}}{a^3}+\frac{\pi }{48}\frac{c_{2,D}+c_{2,N}}{a^2}+O(a^0)\\
=&\frac{7\pi^2 }{1920 a^4}A(\Omega) +\frac{\pi }{48}\frac{2\chi-1}{a^2}+O(a^0).
\end{split}\end{equation}In particular, for small plate separation, the leading term of the Casimir  force is $$\frac{7\pi^2 }{1920 a^4}=\frac{7}{8}\frac{\pi^2 }{240 a^4}A(\Omega),$$ which is $-7/8$ times the Casimir force acting on a pair of perfectly conducting infinite parallel plates, in consistent with the result of Boyer \cite{22} for parallel plates. Notice that this leasing term only depends on the area but not the geometry of $\Omega$.

In the case of rectangular piston, we have the following explicit formula for the zero temperature Casimir force in the $L_1\rightarrow \infty$ limit:
\begin{equation*}
\begin{split}
  F_{\text{Cas}}^{ \infty, T=0}(a; L_2, L_3) =& \frac{7\pi^2 }{1920}\frac{L_2L_3}{a^4}-\frac{\pi }{96}\frac{1}{a^2}-\frac{\pi^2}{720}\frac{L_2}{L_3^3}-\frac{\zeta_R(3)}{16\pi L_3^2} -\frac{1}{2L_2^{\frac{3}{2}}L_3^{\frac{1}{2}}}\sum_{k_2=1}^{\infty}\sum_{k_3=1}^{\infty}\left(\frac{k_2}{k_3}\right)^{\frac{3}{2}}\\&\times K_{\frac{3}{2}}\left(\frac{2\pi k_2k_3 L_3}{L_2}\right) +\frac{\pi L_2 L_3}{4a^3}\sum_{k_1=0}^{\infty}\sum_{(k_2, k_3)\in \mathbb{Z}^2\setminus\{\mathbf{0}\}}\frac{\left(k_1+\frac{1}{2}\right)^2}{\sqrt{(k_2L_2)^2+(k_3L_3)^2}}\\
 &\times \exp\left(-\frac{\pi(2k_1+1)}{a}\sqrt{(k_2 L_2)^2+(k_3L_3)^2}\right).
\end{split}
\end{equation*}In particular,
\begin{equation*}\begin{split} F_{\text{Cas}}^{ \infty, T=0}(a)=& \frac{7\pi^2 }{1920}\frac{L_2L_3}{a^4}-\frac{\pi }{96}\frac{1}{a^2}+O(a^0),
\end{split}\end{equation*}in agreement with the general result \eqref{eq10_6_2}.

\section{Numerical results and Discussions}\label{sec4}
\begin{table}\caption{\label{Tab1} The Casimir force $F_{\text{Cas}} $ and the contribution from   $F_{\text{Cas}}^{  \parallel}$ when $L_2=L_3=0.3\text{m}$ and $T=0\text{K}$.}

\begin{tabular}{|c|c|c|}
\hline
$a$ (m) & $F_{\text{Cas}} $ (N)   & $F_{\text{Cas}}^{   \parallel}$ (N)   \\
\hline
$ 10^{-8} $ & $1.0238\times 10^4$ & $1.0238\times 10^4$\\
$ 10^{-4} $ & $1.0238\times 10^{-12}$ & $1.0238\times 10^{-12}$\\
$ 0.1  $ & $9.0759\times 10^{-25}$ & $1.0238\times 10^{-24}$\\
$ 0.3 $ & $2.6683\times 10^{-27}$ & $1.2640\times 10^{-26}$\\
\hline
\end{tabular}
 \end{table}We compute the magnitude of the Casimir force when the cross section of the piston is a square of dimension 0.3m $\times$ 0.3m, $L_1\rightarrow \infty$ and the temperature is equal to $0\text{K},  1\text{K}, 300\text{K}$ respectively. The results are compared to the contribution from the term $$F_{\text{Cas}}^{  \parallel}=\frac{7\pi^2 }{1920}\frac{L_2L_3}{a^4},$$  the classical term \eqref{eq10_8_6} and the term
$$F_{\text{Cas}}^{\text{classical}, \parallel}=\frac{3\zeta_R(3)}{16 \pi a^3}L_2L_3T, $$ and are tabulated in Tables \ref{Tab1}, \ref{Tab2} and \ref{Tab3}. Notice that when $a/L_2$ is small and $aT  k_B/(\hbar c) \ll 1$, the Casimir force is dominated by the term $F_{\text{Cas}}^{  \parallel}$, whereas if $a/L_2$ is small but $aT  k_B/(\hbar c) \gg 1$, the Casimir force is dominated by the term $F_{\text{Cas}}^{\text{classical}, \parallel}$. When $a/L_2 \simeq 1$, there is a considerable deviation of the Casimir force from the leading terms $F_{\text{Cas}}^{\parallel}$ and $F_{\text{Cas}}^{\text{classical}, \parallel}$. At visible length, the Casimir force is negligible. However, when the separation between the piston and an opposite wall shrinks to 10nm, then the Casimir pressure on the piston is approximately equal to 1 atmospheric pressure at both zero and room temperature.

\begin{table}
\caption{\label{Tab2} The Casimir force $F_{\text{Cas}}$ and the contribution from     the classical term $F_{\text{Cas}}^{  \text{classical}}$ and $F_{\text{Cas}}^{\text{classical}, \parallel}$ when $L_2=L_3=0.3\text{m}$ and $T=1\text{K}$.}

\begin{tabular}{|c|c|c|c|}
\hline
$a$ (m) & $F_{\text{Cas}} $ (N) &   $F_{\text{Cas}}^{\text{classical}}$ (N) & $F_{\text{Cas}}^{\text{classical}, \parallel}$ (N)\\
\hline
$10^{-8}$ & $1.0238\times 10^4$ & $8.9146\times 10^{-2}$ & $8.9146\times 10^{-2}$\\
$10^{-4}$ & $1.0238\times 10^{-12}$ & $8.9146\times 10^{-14}$ & $8.9146\times 10^{-14}$\\
$0.1 $ & $8.1004\times 10^{-23}$ & $8.1004\times 10^{-23} $ & $8.9146\times 10^{-23}$\\
$0.3$ & $5.9862\times 10^{-25}$ & $5.9862\times 10^{-25}$ & $3.3017\times 10^{-24}$\\
\hline
\end{tabular}

\end{table}

\begin{table}
\caption{\label{Tab3} The Casimir force $F_{\text{Cas}}$ and the contribution from     the classical term $F_{\text{Cas}}^{  \text{classical}}$ and $F_{\text{Cas}}^{\text{classical}, \parallel}$ when $L_2=L_3=0.3\text{m}$ and $T=300\text{K}$.}

\begin{tabular}{|c|c|c|c|}
\hline
$a$ (m) & $F_{\text{Cas}} $ (N) &   $F_{\text{Cas}}^{\text{classical}}$ (N) & $F_{\text{Cas}}^{\text{classical}, \parallel}$ (N) \\
\hline
$10^{-8}$ & $1.0238\times 10^4$ & $2.6744\times 10$ & $2.6744\times 10$\\
$10^{-4}$ & $2.6744\times 10^{-11}$ & $2.6744\times 10^{-11}$ & $2.6744\times 10^{-11}$\\
$0.1 $ & $2.4301\times 10^{-20}$ & $2.4301\times 10^{-20} $ & $2.6744\times 10^{-20}$\\
$0.3$ & $1.7959\times 10^{-22}$ & $1.7959\times 10^{-22}$ & $9.9051\times 10^{-22}$\\
\hline
\end{tabular}

\end{table}

For possible applications to nanotechnology, we also compute the Casimir pressure acting on the piston when the cross section of the piston has dimension 100nm $\times$ 100nm and the temperature is 0K and $300$K. The results are tabulated in Tables \ref{Tab4} and \ref{Tab5}. At this length scale, the Casimir force is not very much affected when the temperature change from 0K to 300K. We also find that the Casimir pressure is almost the same as the atmospheric pressure when $a=10\text{nm}$. Reducing $a$ at a rate $r$ will result in the increase of the pressure at the rate $r^4$. Since the Casimir force is pushing the piston outward, this will prevent the undesirable collapse of the piston to the opposite wall.

\begin{table}\caption{\label{Tab4} The Casimir pressure $P_{\text{Cas}} $ and the contribution from   $P_{\text{Cas}}^{  \parallel}$ when $L_2=L_3=100\text{nm}$ and $T=0\text{K}$.}

\begin{tabular}{|c|c|c|}
\hline
$a$ (nm) & $P_{\text{Cas}} $ (N/m$^2$)   & $P_{\text{Cas}}^{   \parallel}$ (N/m$^2$)   \\
\hline
$ 1 $ & $1.1375\times 10^9$ & $1.1376\times 10^9$\\
$ 10 $ & $1.1271\times 10^{5}$ & $1.1376\times 10^{5}$\\
$ 50  $ & $1.3248\times 10^{2}$ & $1.8202\times 10^{2}$\\
$ 100 $ & $2.4015$ & $1.1376\times 10$\\
\hline
\end{tabular}
 \end{table}

\begin{table}
\caption{\label{Tab5} The Casimir pressure $P_{\text{Cas}}$ and the contribution from     the classical term $P_{\text{Cas}}^{\text{classical}}$ and $P_{\text{Cas}}^{\text{classical}, \parallel}$ when $L_2=L_3=100\text{nm}$ and $T=300\text{K}$.}

\begin{tabular}{|c|c|c|c|}
\hline
$a$ (nm) & $P_{\text{Cas}} $ (N/m$^2$) &   $P_{\text{Cas}}^{\text{classical}}$ (N/m$^2$) & $P_{\text{Cas}}^{\text{classical}, \parallel}$ (N/m$^2$) \\
\hline
$1$ & $1.1375\times 10^9$ & $2.9715\times 10^5$ & $2.9715\times 10^5$\\
$10$ & $1.1273\times 10^{5}$ & $2.9641\times 10^{2}$ & $2.9715\times 10^{2}$\\
$50 $ & $1.4299\times 10^{2}$ & $1.7344 $ & $2.3772$\\
$100$ & $1.2061\times 10$ & $5.3876\times 10^{-2}$ & $2.9715\times 10^{-1}$\\
\hline
\end{tabular}

\end{table}

\begin{figure}\centering \epsfxsize=.55\linewidth
\epsffile{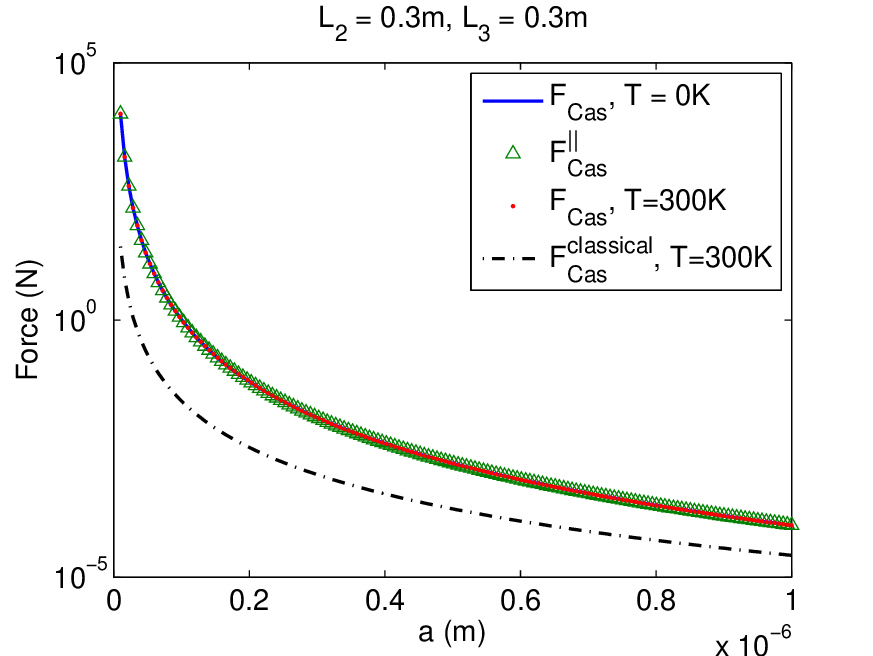} \\ \epsfxsize=.55\linewidth
\epsffile{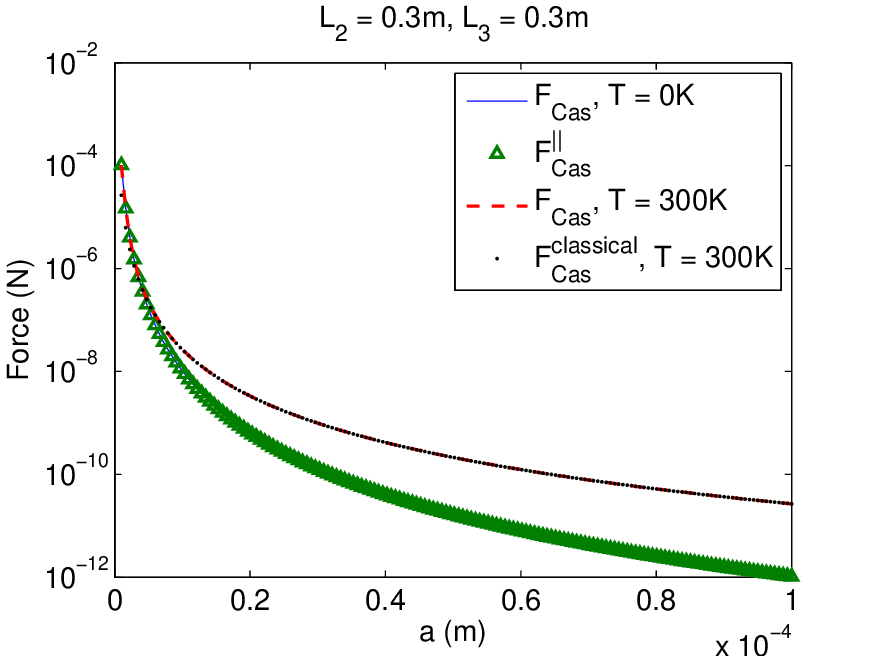}\\ \epsfxsize=.55\linewidth
\epsffile{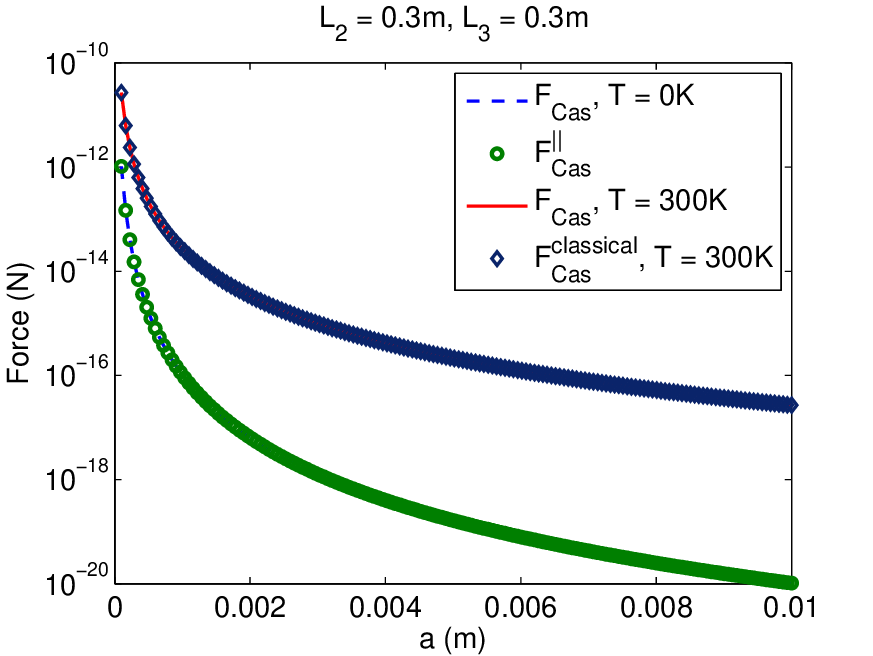} \\ \epsfxsize=.55\linewidth
\epsffile{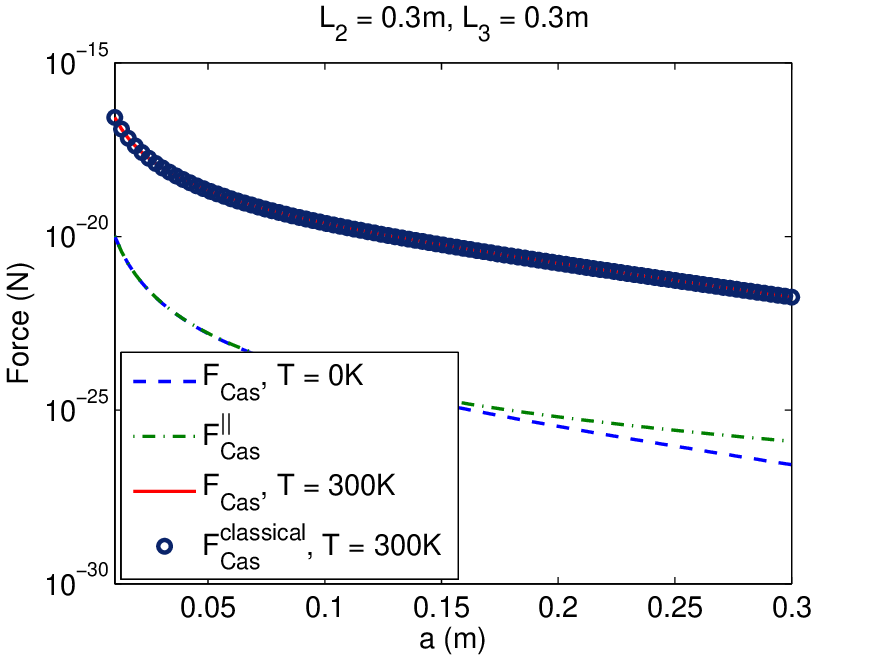}  \caption{\label{f2}The Casimir force $F_{\text{Cas}}$ at $T=0$K and $T=300$K and the contribution from $F_{\text{Cas}}^{\parallel}$ and $F_{\text{Cas}}^{\text{classical}}$ plotted as a function of $a$.}\end{figure}

\begin{figure}\centering \epsfxsize=.55\linewidth
\epsffile{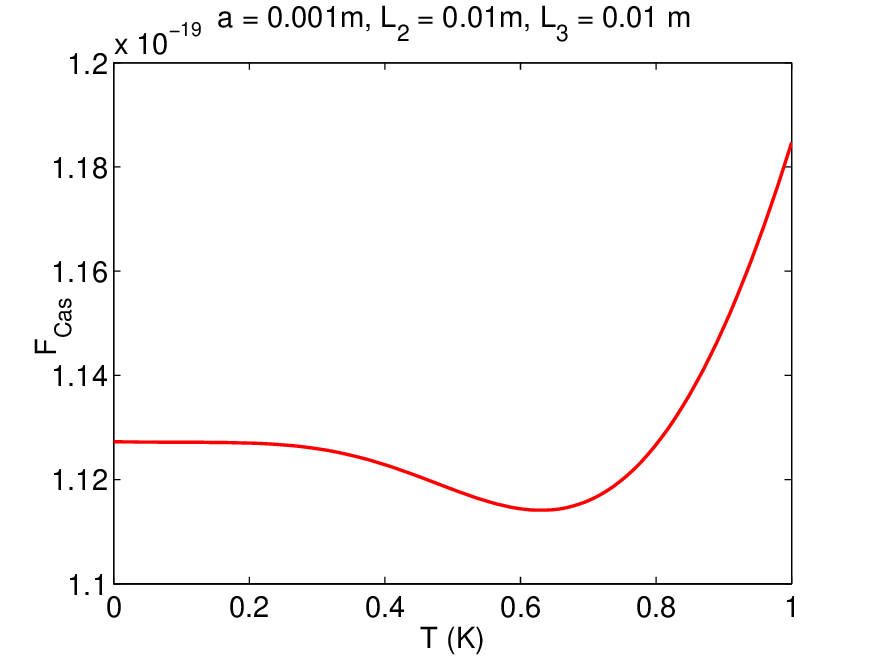}  \caption{\label{f3}This graph shows that the Casimir force is not an increasing function of temperature. Here $a = 0.001$m, $L_2=L_3=$0.01m.}\end{figure}

FIG \ref{f2} shows the behavior of the Casimir force $F_{\text{Cas}}$ as a function of $a$ at $T=0$K and $T=300$K. It is compared to $F^{\parallel}_{\text{Cas}}$ and $F_{\text{Cas}}^{\text{classical}}$. We see that when $a$ is in the range $10$nm to $1\mu$m, the Casimir force at both $0$K and $300$K are dominated by the zero temperature parallel plate term $F_{\text{Cas}}^{\parallel}$. When $a$ is in the range $1\mu$m to $0.3$m, the force is then dominated by the classical term at $T=300$K. From the graphs, one would tend to conclude that the Casimir force is an increasing function of the temperature. However, this is not the case as shown by FIG \ref{f3}.

\section{Conclusion}\label{sec5}

We have shown that for a perfectly conducting piston moving freely inside a cylinder with infinitely permeable walls, the Casimir force acting on the piston is a repulsive force which tends to push the piston to its equilibrium position. At zero temperature, when the separation $a$ between the piston and one of its opposite walls is small, then the magnitude of the Casimir pressure is asymptotically equal to $$\frac{7}{8}\frac{\pi^2 \hbar c }{240 a^4},$$ which is the result obtained by Boyer \cite{22} for a pair of infinite parallel plates, one being infinitely conducting and the other being infinitely permeable. It is $-7/8$ times the zero temperature Casimir pressure acting on a pair of perfectly conducting parallel plates. However, at high temperature, the Casimir pressure is dominated by
 \begin{equation}\label{eq10_6_4}\frac{3}{4} \frac{\zeta_R(3)}{4 \pi a^3} k_B T\end{equation}when $a$ is small. This is $-3/4$ times the Casimir pressure when both plates are perfectly conducting. It is interesting to notice the change from the ratio $7/8$ at zero temperature to the ratio $3/4$ at high temperature.
We   have also shown that   at high temperature, the Casimir force acting on the piston is dominated by a term linear in $T$ known as the classical term. Moreover, when $a$ is small, the classical term is asymptotically equal to the area of the cross section $A(\Omega)$ multiply with the pressure \eqref{eq10_6_4}, which is of order $a^{-3}$. The correction term is of order $ a^0$.

For simplicity, in this paper we have assumed that the piston is perfectly conducting whereas the surrounding walls are infinitely permeable. We would like to consider in a future work the more general case where the piston and its surrounding walls are allowed to have different electric permittivity and magnetic permeability. It would then be interesting to determine the range of the parameters for which the Casimir force acting on the piston is repulsive. One can anticipate that this would have important applications in nanotechnology.

\appendix
\section{Computations of the Casimir energy and asymptotic behavior of the Casimir force}\label{app}

\subsection{Computation of the Casimir energy}
Define the following zeta functions
\begin{equation*}\begin{split}
\zeta_{\Omega, D}(s)= &\sum_{j=1}^{\infty} \omega_{D,j}^{-2s}, \hspace{0.5cm}\zeta_{\text{cylinder, TE}}(s)=\sum_{k=0}^{\infty}\sum_{j=1}^{\infty}\omega_{\text{TE},j,k}^{-2s},\\
\zeta_{\text{TE}}(s) =&\sum_{k=0}^{\infty}\sum_{j=1}^{\infty}\sum_{n=-\infty}^{\infty}\left(\omega_{\text{TE},j,k}^2+[2\pi T]^2\right)^{-s},\end{split}
\end{equation*}and the   heat kernels
\begin{equation*}\begin{split}
K_{\Omega, D}(t) =&\sum_{j=1}^{\infty} e^{-t\omega_{D,j}^2},\\
K_{\text{cylinder}, \text{TE}}(t)=&\sum_{k=0}^{\infty}\sum_{j=1}^{\infty}e^{-t\omega_{TE,j,k}^2}=\sum_{k=0}^{\infty} e^{-t\left[\frac{\pi\left(k+\frac{1}{2}\right)}{L}\right]^2}K_{\Omega,D}(t).
\end{split}\end{equation*}The zeta functions $\zeta_{\Omega,N}(s)$, $\zeta_{\text{cylinder}, \text{TM}}(s)$, $\zeta_{ \text{TM}}(s)$ and the heat kernels $K_{\Omega, N}(t)$ and $K_{\text{cylinder}, \text{TM}}(s)$ for the TM modes are defined analogously. It is well known that the heat kernel $K_{\Omega, D}(t)$ has the following asymptotic expansion
\begin{equation*}
K_{\Omega, D}(t)\simeq \sum_{l=0}^{M} c_{D,l} t^{\frac{l-2}{2}} +O\left(t^{\frac{M-1}{2}}\right), \;\;\text{as}\;t\rightarrow 0^+.
\end{equation*}Therefore,
\begin{equation*}\begin{split}
K_{\text{cylinder}, \text{TE}}(t)=&\frac{L}{2\sqrt{\pi t}}\left(1+2\sum_{k=1}^{\infty}(-1)^ke^{-\frac{k^2L^2}{t}}\right)K_{\Omega, D}(t)\\
\simeq & \frac{L}{2\sqrt{\pi }}\sum_{l=0}^{M} c_{D, l} t^{\frac{l-3}{2}} +O\left(t^{\frac{M-2}{2}}\right), \;\;\text{as}\;t\rightarrow 0^+.
\end{split}
\end{equation*}Consequently, the   function $\Gamma(s)\zeta_{\text{cylinder, TE}}(s)$ has at most simple poles at $s=\frac{3-l}{2}$, $l=0, 1, 2, \ldots$ with residues
\begin{equation*}
\text{Res}_{s=\frac{3-l}{2}}\Bigl\{\Gamma(s)\zeta_{\text{cylinder, TE}}(s)\Bigr\}=\frac{L}{2\sqrt{\pi}}c_{D,l}.
\end{equation*}

The $\lambda\rightarrow 0^+$ behavior of the $T=0$ part of the cut-off dependent Casimir energy \eqref{eq10_3_1} can   be determined as follows:
\begin{equation*}
\begin{split}
E_{\text{Cas, TE}}^{\text{cylinder}, T=0}(L)=&\frac{1}{2}\sum_{k=0}^{\infty}\sum_{j=1}^{\infty} \omega_{\text{TE},j,k} e^{-\lambda \omega_{\text{TE},j,k}}\\
=&-\frac{1}{2}\frac{\pa}{\pa\lambda}\frac{1}{2\pi i}\int_{u-i\infty}^{u+i\infty} dz \Gamma(z)\lambda^{-z} \zeta_{\text{cylinder, TE}}\left(\frac{z}{2}\right)\\
\simeq &\frac{6L}{\pi}c_{D,0}\lambda^{-4}+\frac{L}{\sqrt{\pi}}c_{D,1}\lambda^{-3}+\frac{L}{2\pi}c_{D,2}\lambda^{-2}\\&+\frac{L}{4\pi}\left(\log\lambda +\gamma\right)c_{D,4}+\frac{1}{2}\text{FP}_{s=-\frac{1}{2}}\zeta_{\text{cylinder, TE}}(s).
\end{split}
\end{equation*}Here $\gamma$ is the Euler constant and $\text{FP}_{s=-\frac{1}{2}}\zeta_{\text{cylinder, TE}}(s)$ is the finite part of the zeta function $\zeta_{\text{cylinder, TE}}(s)$ at $s=-1/2$. Since
\begin{equation*}\begin{split}\zeta_{\text{cylinder, TE}}(s)=\frac{1}{\Gamma(s)}\int_0^{\infty} t^{s-1}K_{\text{cylinder},\text{TE}}(t)dt,
\end{split}
\end{equation*}a straightforward computation gives
\begin{equation*}
\begin{split}
\frac{1}{2}\text{FP}_{s=-\frac{1}{2}}\zeta_{\text{cylinder, TE}}(s)=&\frac{L}{8\pi}\Bigl\{c_{D, 4}\left[2-\gamma-2\log2\right]-\text{FP}_{s=-1}\left(\Gamma(s)\zeta_{\Omega,D}(s)\right)\Bigr\}\\&
 -\frac{1}{2\pi}\sum_{k=1}^{\infty}(-1)^{k}\sum_{j=1}^{\infty}\frac{\omega_{D,j}}{k}K_1\left(2kL\omega_{D,j}\right),
\end{split}\end{equation*}where $K_{\nu}(z)$ is the modified Bessel function.  Gathering the above results, we find that the  contribution to the cut-off dependent zero temperature Casimir energy of the cylinder $[0, L]\times\Omega$ from the TE modes is given by
\begin{equation}\label{eq9_26_1_a}
\begin{split}
E_{\text{Cas, TE}}^{\text{cylinder}, T=0}(L)=&\frac{6L}{\pi}c_{D,0}\lambda^{-4}+\frac{L}{\sqrt{\pi}}c_{D,1}\lambda^{-3}+\frac{L}{2\pi}c_{D,2}\lambda^{-2}+\frac{L}{8\pi}\left(2\log\lambda +\gamma+2-2\log 2\right)c_{D,4}\\&-\frac{L}{8\pi}\text{FP}_{s=-1}\left(\Gamma(s)\zeta_{\Omega,D}(s)\right)-\frac{1}{2\pi}\sum_{k=1}^{\infty}(-1)^{k}\sum_{j=1}^{\infty}\frac{\omega_{D,j}}{k}K_1\left(2kL\omega_{D,j}\right).
\end{split}
\end{equation}

For the finite temperature Casimir energy, we can use the following formulas. On the   one hand, we have (see e.g. \cite{27}):
\begin{equation}\label{eq9_26_4_a}
\begin{split}
&-\frac{T}{2}\zeta_{\text{TE}}'(0)=\frac{1}{2}\text{FP}_{s=-\frac{1}{2}}\zeta_{\text{cylinder, TE}}(s) -\frac{L}{4\pi}c_{D,4}\left(1-\log 2\right) +T\sum_{k=0}^{\infty}\sum_{j=1}^{\infty} \log\left( 1- e^{-\omega_{\text{TE}, j,k}/T}\right).
\end{split}
\end{equation}On the other hand,
\begin{equation}\label{eq9_26_5_a}
\begin{split}
-\frac{T}{2}\zeta_{\text{TE}}'(0)=&-\frac{L}{8\pi}\Bigl\{\gamma c_{D,4}+\text{FP}_{s=-1}\left(\Gamma(s)\zeta_{\Omega,D}(s)\right)\Bigr\}
-\frac{LT}{\pi}\sum_{l=1}^{\infty}\sum_{j=1}^{\infty} \frac{\omega_{D,j}}{l}K_1\left(\frac{l\omega_{D,j}}{T}\right)\\&+\frac{T}{2}\sum_{l=-\infty}^{\infty}\sum_{j=1}^{\infty} \log\left(1+e^{-2L\sqrt{(2\pi lT)^2+\omega_{D,j}^2}}\right).
\end{split}
\end{equation}From these, we find that the cut-off dependent Casimir energy due to TE modes is given by
\begin{equation}\label{eq9_26_2_a}
\begin{split}
E_{\text{Cas, TE}}^{\text{cylinder}}(L)=&\frac{6L}{\pi}c_{D,0}\lambda^{-4}+\frac{L}{\sqrt{\pi}}c_{D,1}\lambda^{-3}+\frac{L}{2\pi}c_{D,2}\lambda^{-2}
\frac{L}{8\pi}\left(2\log\lambda  +2+\gamma-2\log 2\right)c_{D,4} \\&-\frac{L}{8\pi} \text{FP}_{s=-1}\left(\Gamma(s)\zeta_{\Omega,D}(s)\right)-\frac{LT}{\pi}\sum_{l=1}^{\infty}\sum_{j=1}^{\infty} \frac{\omega_{D,j}}{l}K_1\left(\frac{l\omega_{D,j}}{T}\right)\\&+\frac{T}{2}\sum_{l=-\infty}^{\infty}\sum_{j=1}^{\infty} \log\left(1+e^{-2L\sqrt{(2\pi lT)^2+\omega_{D,j}^2}}\right).
\end{split}
\end{equation}

\subsection{The leading behavior of the classical term of the Casimir force at small plate separation}

Here we compute the small-$a$ asymptotic behavior of the classical term of the Casimir force \eqref{eq10_8_6}. We have
\begin{equation*}\begin{split}
F_{\text{Cas}}^{\text{classical}}(a; L_1)=&T\sum_{ \omega_{D,j}, \omega_{N,j}}  \frac{\omega}{e^{2a\omega}+1}+O(a^0)\\
=&-\frac{T}{\sqrt{\pi}} \sum_{k=1}^{\infty}\sum_{  \omega } (-1)^k \omega^2  \int_0^{\infty} t^{-\frac{3}{2}}\exp\left(-t(ka)^2-\frac{ \omega^2 }{t}\right)dt\\=&\frac{T}{\sqrt{\pi}}\frac{1}{2\pi i}\int_{3-i\infty}^{3+i\infty}  a^{1-2z}\Gamma\left(z-\frac{1}{2}\right) \zeta_R(2z-1) \left(1-2^{2-2z}\right)\\&\hspace{3cm}\times \Gamma(z)\left(\zeta_D(z-1)+\zeta_N(z-1)\right)dz\\
=&T\left\{ \frac{3\zeta_R(3)}{8a^3}\left(c_{0,D} +c_{0,N}\right)+\frac{\pi^{\frac{3}{2}}}{24 a^2}\left(c_{1,D}+c_{1,N}\right)\right\}+O(a^0)\\
=&\frac{3\zeta_R(3)}{16 \pi a^3}A(\Omega)T+O(a^0).\end{split}
\end{equation*}

\subsection{Verification of \eqref{eq10_3_2}}
\begin{equation}\label{eq10_3_2_a}\begin{split}
\frac{T}{\pi}\sum_{l=1}^{\infty}\sum_{j=1}^{\infty} \frac{\omega_{D,j}}{l}K_1\left(\frac{l\omega_{D,j}}{T}\right)=&\frac{1}{4\pi}\int_0^{\infty} \sum_{l=1}^{\infty}\sum_{j=1}^{\infty} \exp\left\{-\frac{tl^2}{4T^2}-\frac{\omega_{D,j}^2}{t}\right\}dt\\
=&\frac{1}{8\pi^2 i}\int_{2-i\infty}^{2+i\infty} \Gamma(z+1)\zeta_R(2z+2)(2T)^{2z+2}\Gamma(z)\zeta_{\Omega, D}(z)dz\\
=&\frac{2\pi^3}{45}c_{0,D}T^4+\frac{\zeta_R(3)}{\sqrt{\pi}}c_{1,D}T^3+\frac{\pi}{6}c_{2,D}T^2+O(T).
\end{split}\end{equation}

\section{Alternative formulas for the Casimir force acting on a rectangular piston}\label{app2}
Here we present two exact formulas for the Casimir force when the piston has rectangular cross section. These formulas are useful when the temperature $T$ is high and the separation $a$ is small. The formulas can be used to study the behavior of the Casimir force when the combination $aT$ is small and large respectively. They can be derived using the Chowla--Selberg formula as presented in \cite{18}.

\subsection{$aT\ll 1$} \begin{equation*}\begin{split}
&F_{\text{Cas}}^{\infty}(a; L_2, L_3) =\frac{7}{8}\frac{\pi^2 L_2L_3}{240 a^4}-\frac{\pi}{96a^2}-\frac{L_3T}{8\pi L_2^2}\zeta_R(3)+\frac{\pi T^2}{12}-\frac{\pi}{24}\frac{T}{L_3}-\frac{\pi^2 T^4 L_2L_3}{45} -\frac{T}{L_2}\\&\times\sum_{k_2=1}^{\infty}\sum_{k_3=1}^{\infty}\frac{k_2}{k_3}K_1\left(\frac{2\pi k_2 k_3 L_3}{L_2}\right) -\frac{\pi}{2a^2}\sum_{k_1=0}^{\infty}\frac{k_1+\frac{1}{2}}{\exp\left(\frac{\pi\left(k_1+\frac{1}{2}\right) }{aT}\right)-1} -\frac{\pi T L_2 L_3}{a^3}\sum_{k_1=0}^{\infty}\left(k_1+\frac{1}{2}\right)^2\\&\times\log\left(1-\exp\left(-\frac{\pi\left(k_1+\frac{1}{2}\right)}{aT}\right)\right)
+\frac{\pi L_2L_3 T}{ a^3} \sum_{(k_2, k_3)\in \widehat{\mathbb{Z}^2}}\sum_{k_1=0}^{\infty}\sum_{l=-\infty}^{\infty}\left(k_1+\frac{1}{2}\right)^2
\\&\times  K_0\left(2\pi\sqrt{\left(\left[\frac{k_1+\frac{1}{2}}{a}\right]^2+[2lT]^2\right)\left([k_2L_2]^2+[k_3L_3]^2\right)}\right).
\end{split}\end{equation*}
 \subsection{$aT\gg 1$}

 \begin{equation*}\begin{split}
& F_{\text{Cas}}^{\infty}(a; L_2, L_3) =\frac{3\zeta_R(3) T}{16 \pi a^3}L_2L_3 -\frac{L_3T}{8\pi L_2^2}\zeta_R(3)-\frac{\pi}{24}\frac{T}{L_3}-\frac{T}{L_2}\sum_{k_2=1}^{\infty}\sum_{k_3=1}^{\infty}\frac{k_2}{k_3}K_1\left(\frac{2\pi k_2 k_3 L_3}{L_2}\right)\\&-\frac{2T^2 L_2L_3}{a^2}\sum_{k=1}^{\infty}\frac{(-1)^{k}}{k^2} \frac{e^{4\pi k aT}}{\left(e^{4\pi k aT}-1\right)^2}-\frac{L_2L_3 T}{2\pi a^3}\sum_{k=1}^{\infty}\frac{(-1)^k}{k^3}\frac{1}{e^{4\pi k aT}-1}+\frac{4\pi T^3L_2L_3}{a}\sum_{l=1}^{\infty}l^2\\&\times\log\left(1+e^{-4\pi l aT}\right)-2\pi T^2  \sum_{l=1}^{\infty}\frac{l}{e^{4\pi l aT}+1}
 +\frac{\pi L_2L_3 T}{ a^3} \sum_{(k_2, k_3)\in \widehat{\mathbb{Z}^2}}\sum_{k_1=0}^{\infty}\sum_{l=-\infty}^{\infty} \left(k_1+\frac{1}{2}\right)^2
\\&\times K_0\left(2\pi\sqrt{\left(\left[\frac{k_1+\frac{1}{2}}{a}\right]^2+[2lT]^2\right)\left([k_2L_2]^2+[k_3L_3]^2\right)}\right).
 \end{split}\end{equation*}

\vspace{1cm}\noindent
\textbf{Aknowledgment}
This project is   supported by Ministry of Science, Technology and Innovation, Malaysia under SAGA fund P96c and e-Science fund 06-02-01-SF0080.


\begin{thebibliography}{10}

\bibitem{1}
H.~B.~G. Casimir, \emph{On the attraction between two perfectly
conducting plates}, Proc. Kon. Nederland. Akad. Wetensch. \textbf{B51} (1948), 793--795.
\bibitem{2}
Jan Ambj{\o}rn and S.~Wolfram, \emph{Properties of the vacuum. {I}.
  {M}echanical and thermodynamic}, Ann. Physics \textbf{147} (1983), 1--32.
\bibitem{3}
Steven~K. Blau, Matt Visser, and Andreas Wipf, \emph{Zeta functions
and the {C}asimir energy}, Nuclear Phys. B \textbf{310} (1988), 163--180.

\bibitem{4}R. L. Jaffe, \emph{Unnatural Acts: Unphysical Consequences of
Imposing Boundary Conditions on Quantum Fields}, in Quantum Field
Theory Under the Influence of External Conditions, edited by K.
Milton, (Rinton Press, Paramus, NJ, 2004).

\bibitem{nn1} B. Geyer, G. L. Klimchitskaya and  V. M. Mostepanenko, \emph{Thermal Casimir effect in ideal metal rectangular boxes}, preprint arXiv:0808.3754, to appear in Euro. Phys. J. C.

\bibitem{nn2} S. A. Fulling, L. Kaplan, K. Kirsten, Z. H. Liu and K. A. Milton, \emph{ Vacuum Stress and Closed Paths in Rectangles, Pistons, and Pistols}, preprint arXiv:0806.2468.

\bibitem{5}
R. M. Cavalcanti, \emph{Casimir force on a piston}, Phys. Rev. D
\textbf{69} (2004), 065015.

\bibitem{6}
M. P. Hertzberg, R. L.  Jaffe, M. Kardar, A. Scardicchio,
\emph{Attractive Casimir forces in a closed geometry}, Phys. Rev.
Lett. \textbf{95} (2005), 250402.


\bibitem{7}
M. P. Hertzberg, R. L.  Jaffe, M. Kardar, A. Scardicchio,
\emph{Casimir forces in a piston geometry at zero and finite
temperatures}, Phys. Rev. D \textbf{76} (2007), 045016.

\bibitem{8}
V. N. Marachevsky, \emph{One loop boundary effects: techniques and
applications}, preprint arXiv: hep-th/0512221 (2005).

\bibitem{9}
G. Barton, \emph{Casimir piston and cylinder, perturbatively}, Phys.
Rev. D \textbf{73} (2006), 065018.

\bibitem{10}
V. N. Marachevsky, \emph{Casimir energy of two plates inside a
cylinder}, Phys. Rev. D \textbf{75} (2007), 085019.

\bibitem{11}
A. Edery, \emph{Casimir piston for massless scalar fields in three
dimensions}, Phys. Rev. D \textbf{75} (2007), 105012.

\bibitem{12}
A. Edery and I. Macdonald, \emph{Cancellation of nonrenormalizable
hypersurface divergences and the d-dimensional Casimir piston}, J.
High Energy Phys. \textbf{9} (2007), 0709:005.

\bibitem{13}X. H. Zhai and X. Z.Li, \emph{
Casimir pistons with hybrid boundary conditions}, Phys. Rev. D
\textbf{76} (2007), 047704.




\bibitem{14}
V. N. Marachevsky, \emph{Casimir interaction: pistons and cavity},
J. Phys. A: Math. and Theor. \textbf{41} (2008), 164007.





\bibitem{15} A. Edery, V. N. Marachevsky, \emph{The perfect magnetic
conductor (PMC) Casimir piston in d+1 dimensions}, Phys. Rev. D \textbf{78}, 025021 (2008).

\bibitem{16}
H. Cheng, \emph{The Casimir force on a piston in the spacetime with
extra compactified dimensions}, Phys. Lett. B \textbf{668}, 72--77.



\bibitem{17}
S. C. Lim and L. P. Teo, \emph{Three dimensional Casimir piston for massive scalar fields}, preprint arXiv: hep-th:0807.3613.




\bibitem{18}
S. C. Lim and L. P. Teo, \emph{Casimir piston at zero and finite temperature}, preprint arXiv: hep-th:0808.0047.

\bibitem{19}
X. H. Zhai, Y. Y. Zhang and X. Z. Li, \emph{Casimir Pistons for Massive Scalar Fields},  preprint arXiv: hep-th:0808.0062.

\bibitem{20}
F. M. Serry, D. Walliser and G. J. Maclay, \emph{The anharmonic
Casimir oscillator (ACO)-the Casimir effect in amodel
microelectromechanical system}, J. Microelectromech. Syst.
\textbf{4} (1995), 193--205.

\bibitem{21} F. M. Serry, D. Walliser and G. J.
Maclay, \emph{The role of the Casimir effect in the static
deflection and stiction of membrane strips in microelectromechanical
systems (MEMS)}, J.   Applied Phys. \textbf{84}, 2501--2506 (1998).


\bibitem{22} T. H. Boyer, \emph{Van der Waals forces and zero-point energy for dielectric and permeable materials}, Phys. Rev. A \textbf{9} (1974), 2078--2084.

\bibitem{23} F. C. Santos, A. Tenorio and A. C. Tort,  \emph{Zeta function method and repulsive Casimir forces for an unusual pair of plates at finite temperature}, Phys. Rev. D \textbf{60} (1999), 105022.

\bibitem{24}  J. C. da Silva, A. M. Neto, H. Q. Placido, M. Revzen M, A. E. Santana, \emph{Casimir effect for conducting and permeable plates at finite temperature}, Physica A \textbf{292} (2001), 411--421.

\bibitem{25} O. Kenneth, I. Klich, A. Mann and M. Revzen, \emph{Repulsive Casimir forces}, Phys. Rev. Lett. \textbf{89} (2002), 033001.

\bibitem{n2} C. G. Shao, D. L. Zheng and J. Luo, \emph{Repulsive Casimir effect between anisotropic dielectric and permeable plates}, Phys. Rev A \textbf{74} (2006), 012103.

\bibitem{n1} S. A. Fulling, L. Kaplan and J. H. Wilson, \emph{Vacuum energy and repulsive Casimir forces in quantum star graphs}, Phys. Rev A \textbf{76} (2007), 012118.




\bibitem{26}H. P. Baltes and E. R. Hilf, \emph{Spectra of finite systems}, (Bibliographisches Institut, Mannheim, 1976).
\bibitem{27}
S.C. Lim and L.P. Teo, \emph{Finite temperature Casimir energy in
closed rectangular cavities: a rigorous derivation based on zeta
function technique}, J. Phys. A: Math. Theor. \textbf{40} (2007),
11645-11674.
 \end{thebibliography}
\end{document}